\documentclass[pre,showpacs,floatfix,twocolumn]{revtex4}
\usepackage{graphicx}
\usepackage{amssymb}
\usepackage{dcolumn}
\usepackage{bm}

\usepackage{color}

\begin{document}

\title{Creep dynamics of viscoelastic interfaces}
\author{E. A. Jagla}

\affiliation{Centro At\'omico Bariloche and Instituto Balseiro, Comisi\'on Nacional de Energ\'{\i}a At\'omica, 
(8400) Bariloche, Argentina}

\begin{abstract}

The movement of a purely elastic interface driven on a disordered energy potential is characterized by a depinning transition: when the pulling force $\sigma$ is larger than some critical value $\sigma_1$ the system is in a flowing regime and moves at a finite velocity. 
If $\sigma < \sigma_1$ the interface remains pinned and its velocity is zero.  
We show that for a one-dimensional interface, the inclusion of viscoelastic relaxation produces the appearance of an intervening regime between the pinned and the flowing phases in a well defined stress interval $\sigma_0<\sigma<\sigma_1$, in which the interface evolves through a sequence of avalanches that give rise to a creep process. As
$\sigma \to\sigma_0^+$ the creep velocity vanishes in an universal way that is governed by a directed percolation process. As $\sigma \to\sigma_1^-$ the creep velocity increases as a power law due to the increase of the typical size of the avalanches. The present observations may serve to improve the understanding of fatigue failure mechanisms.

\end{abstract}
\maketitle

\section{Introduction}

An elastic interface driven through a disordered potential is a generic model for different physical systems, as domain walls in ferromagnetic materials\cite{ferro1,ferro2,ferro3}, wetting fronts on a rough substrate\cite{wet1,wet2}, and seismic fault dynamics\cite{eq1,eq2,eq3}. When driving the interface at a small, constant velocity,  the dynamical evolution proceeds through abrupt events called avalanches. 
When the elastic interface is driven at constant external force instead, a depinning transition occurs\cite{depinning}: Below some critical applied stress $\sigma_1$
the interface remains pinned, and the configuration is stationary. Above this threshold, the system does not reach an equilibrium configuration, and the dynamics proceeds continuously in time, in a flowing regime with a finite velocity. This velocity critically vanishes as the applied stress is reduced towards $\sigma_1$.

The existence of a true pinned phase for $\sigma<\sigma_1$ relies on the absence of thermally activated effects. If they are present (i.e., if temperature is different from zero) the energy barriers that prevent the evolution of the system are eventually surmounted, and the system can creep at a finite velocity\cite{creep1,creep2}. The velocities generated by the creep process are much smaller than those of the flowing regime, so the value $\sigma_1$ still signals the transition between a low velocity creep regime and a large velocity flow regime. In the presence of thermally activated processes, the velocity of the interface vanishes only when $\sigma \to 0$.

Experimentally, the creep regime may cause fatigue failure\cite{fatiga1,fracture1,fatiga2}, and is a concern in the performance of mechanical components.
It may induce the failure of components after a prolonged time of service at applied loads well below the nominal fracture strength. This behavior is captured in some phenomenological laws, as for instance the Basquin law\cite{basquin}, that states that the lifetime of a component is proportional to some negative power of the applied load. 
The phenomenology of static fatigue failure typically involves additional features that are not appropriately captured by the thermal creep mechanism alone. In many cases a {\em fatigue limit} exists, such that there is no progression of the damage at all if the applied load is below this limit
\cite{fatiga1}. Theoretical explanations of this fact have relied upon the existence of {\em healing mechanisms} in the material\cite{kun1,kun2,alava}, that compete with creep, generating a fatigue limit at a finite applied load. 
 
In this paper we investigate theoretically an alternative mechanism that can produce the slow advance of an interface on a disordered media, then giving insight into the possible mechanisms of fatigue failure under constant applied loads. It is based on the existence of viscoelastic effects in the material. Even if temperature is set to 0 (in the sense that energy barriers can not be surmounted), we find that the interface advances at a finite (but small) velocity in a well defined range of the applied stress $\sigma$, namely $\sigma_0<\sigma<\sigma_1$.
Below $\sigma_0$ there is no advance at all, then this value represents the {\em fatigue limit} of the material. 
The dynamics near the fatigue limit is universal, and can be described as a kind of contact process\cite{hinri}.
The advance occurs through a  sequence of avalanches, that become progressively larger as $\sigma\to \sigma_1$, where the creep regime crosses over to the flow regime. This crossover becomes a sharp transition in the limit in which the temporal scale of the viscoelastic relaxation is much larger than the scale in which individual avalanches develop. In this last case the average size of the creep avalanches diverges as $\sigma\to  \sigma_1$.

The paper is organized as follows. In Section II we define in detail the viscoelastic model used, and give details of the numerical simulation technique. Results are presented in Section III, whereas discussions and conclusions are contained in Section IV.

\section{Model and simulation technique}

\begin{figure}[h]
\includegraphics[width=6cm,clip=true]{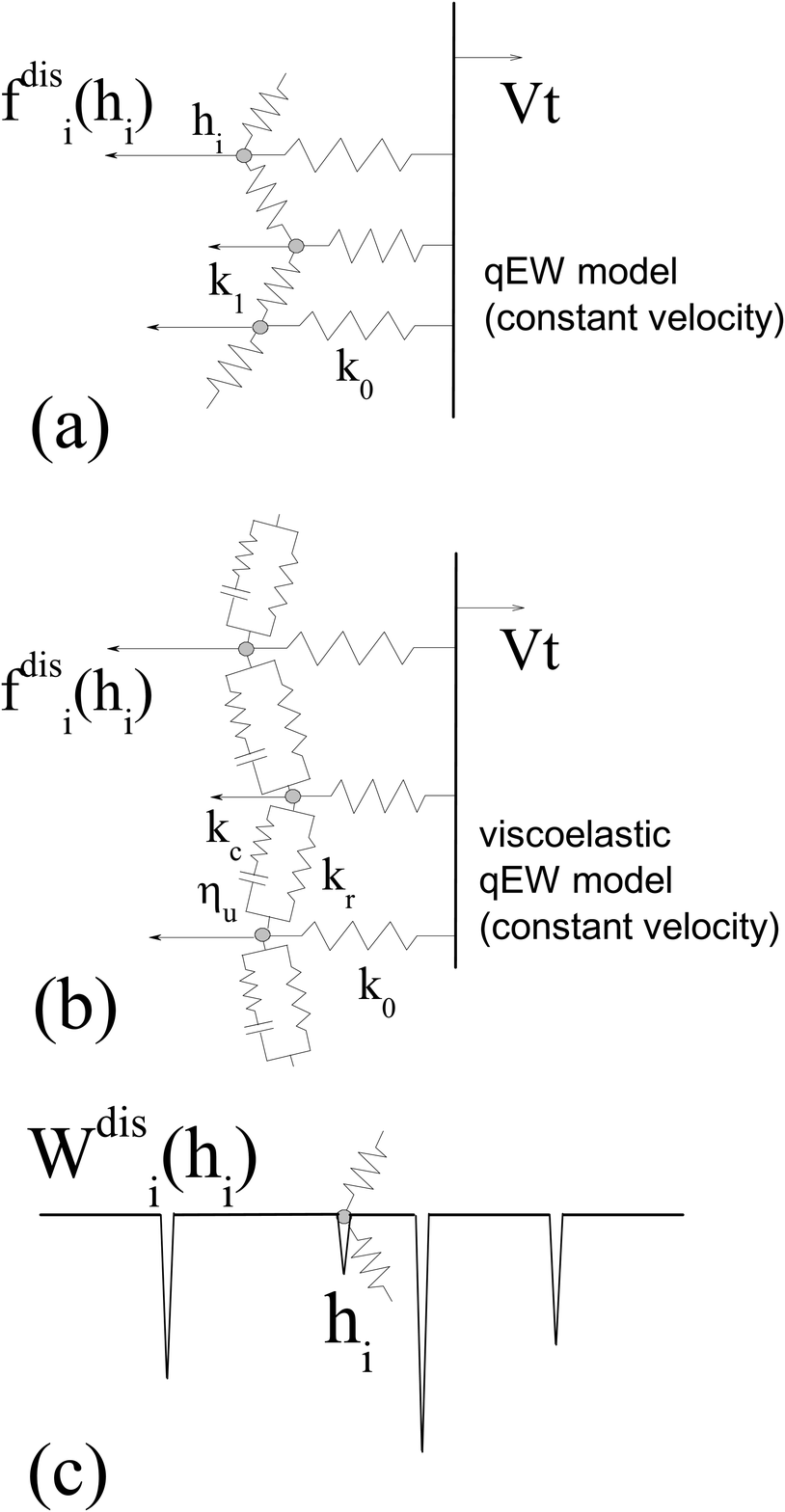}
\caption{(a) Sketch of the discrete quenched Edwards-Wilkinson model, in the constant velocity driving setup. (b) The viscoelastic version discussed here. (c) Sketch of one of the sites in the discrete pinning potential used, from which the pinning forces are obtained: $f_i^{dis}(h_i)=-dW_i^{dis}/dh_i$. The positions of the narrow pinning wells  and their strengths are randomly distributed.
}
\label{f1}
\end{figure}

The growing of a crack on a material under load, can be described by treating the one dimensional crack front as an elastic interface that is driven through a disordered pinning potential, representing the material imperfections at a microscopic scale. An appropriate model that captures the essence of this phenomenon is the quenched Edwards Wilkinson (qEW) model\cite{stanley} that in one spatial dimension and in a discrete description, can be represented by the equation
\begin{equation}
\eta \partial_t h_i =  F_i + f_i^{dis} (h_i)  + k_1 \Delta h_i  
\label{qew}
\end{equation}
The model is schematically represented in Fig. \ref{f1}(a).
$ h_i$ are the dynamical variables of the system defined on discrete sites $i$, $f_i^{dis}$ represents the pinning forces at different sites, and $F_i$ is the driving force on the interface. 
Two driving mechanisms are usually considered for the qEW model. In {\em constant force} driving, the 
value of $F_i$ is constant, and independent of $i$, and represents the stress applied on the system, namely $F_i\equiv \sigma$.
In this case there is a critical stress $\sigma_1$ that separates a pinned regime from a flowing, depinned regime.
In {\em constant velocity} driving $F_i$ is chosen to be of the form $F_i=k_0(Vt-h_i)$, representing a driving at constant velocity $V$ through springs of stiffness $k_0$. The average stress in the system in this case is given by $\sigma=k_0(Vt-\overline h_i)$. 
In the limit of $V \to 0$, the dynamics of the constant velocity case consists of a sequence of avalanches, that have a typical duration that is controlled by the value of $\eta$. We will take formally $\eta\to 0$, and in this sense, the avalanches will be considered as instantaneous.
Constant velocity driving is connected with constant force driving in the limit $k_0\to 0$. In fact, in constant velocity driving the values of $F_i$ for all $i$ tend to $\sigma_1$ as $k_0\to 0$ (see below). 

We work in the case of a discrete pinning potential in which $f_i^{dis}(h_i)$ is different from zero only in some discrete set of values of $h_i$, that represent the positions of pinning centers (Fig. \ref{f1}(c)). Each pinning  center is characterized by the force that is necessary to apply in order to extract a particle from it. These values are noted $f_i^{th}$, and are drawn from a Gaussian distribution with zero mean and unitary variance. The location of the pinning centers along the $h$ direction is random, with a mean separation $z_0=0.1$.

The viscoelastic qEW model introduced in \cite{jlr}, takes into account the possibility of relaxation effects in the material, by  replacing the $k_1$ springs by linear viscoelastic elements, as depicted in Fig. \ref{f1}(b). The equations of the model in the constant force setup are given by:
\begin{eqnarray}
\eta \partial_t h_i
&=&  f_i^{dis}(h_i)+G_i + D_i+\sigma\label{hdot}\\
\eta_u \partial_t D_i +k_c D_i&=& \eta_u k_c (\Delta \partial_t h)_i.
\label{fdot}  
\end{eqnarray}
where $\sigma$ is the externally applied force, 
and $D_i$ and $G_i$ represent the forces onto $h_i$ exerted by the $k_r$ and $k_c$ branches, respectively.
Eq. (\ref{fdot}) describes the relaxation of the force through the $k_c$ branches, due to existence of the dashpot elements, characterized by the constant $\eta_u$.
The value of the dashpot $\eta_u$ sets a new time scale in the system given by $\eta_u/k_c$.
We work in the case in which this time scale is much larger than the typical timescale of individual avalanches, namely $\eta_u\gg\eta$. This simplifies the evaluation of the dynamical evolution, as the $\eta_u$ elements remain blocked during the avalanche. In addition, due to the discrete pinning potential used, the values of $h_i$ remain fixed as viscoelastic elements relax, until the force over any $h_i$ reaches the critical value and a new avalanche is triggered.

The concrete algorithm that is followed to evolve the system in time can be summarized as follows.
For fixed values of $h_i$, the forces $D_i$ relax according to 
\begin{equation}
D_i(t)=D_i(t_0)\exp{\left (-k_ct/\eta_u \right )}
\label{d}
\end{equation}
which is the solution to Eq. (\ref{fdot}) when $h$'s are kept constant.
This relaxation is followed until the force onto some $h_i$ reaches the threshold value: $f_i^{th}=G_i + D_i+\sigma$. At this point, an avalanche starts at position $i$, producing the advance of $h_i$ to the next potential well $h_i \leftarrow h_i +z$, and a corresponding rearrangement of the forces according to:
\begin{eqnarray}
D_i&\leftarrow D_i-2k_cz\\
F_i&\leftarrow F_i-2k_rz\\
D_j&\leftarrow D_j+k_c z\\
F_j&\leftarrow F_j+k_r z
\end{eqnarray}
where $j$ are the two neighbor sites to $i$, and the value of $f_i^{th}$ is renewed from its probability distribution. All successive unstable sites are 
treated in the same way until there are no more unstable sites. Once the avalanche is exhausted, the relaxation of $D_i$ according to Eq. (\ref{d}) is re-initiated until the next instability. For $\sigma$ in the creep regime $\sigma_0<\sigma<\sigma_1$ this dynamics produces an infinite sequence of avalanches, and the interface advances with a finite velocity that can be numerically determined.

\section{Results}

In \cite{jlr}, the properties of the viscoelastic qEW model were studied both in a mean field limit, and in two spatial dimensions, in the constant velocity driving case. It was found that in those cases viscoelasticity induces a non-stationary dynamics in time, that manifests in particular in the fact that the constant velocity driving does not tend to a uniform stress case in the $k_0\to 0$ limit. 

\begin{figure}[h]
\includegraphics[width=8cm,clip=true]{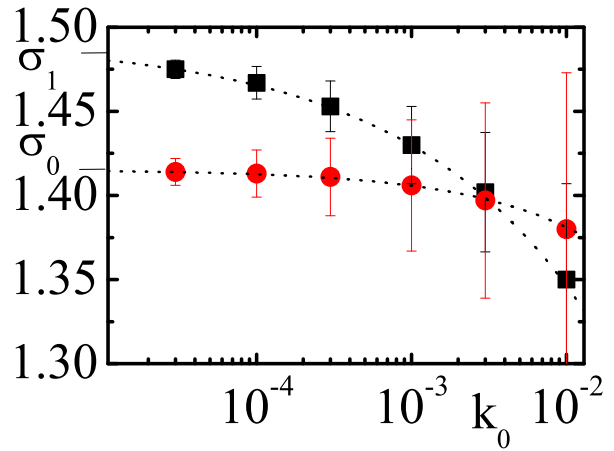}
\caption{The average stress $\sigma$ as a function of $k_0$ in constant velocity driving simulations, for the qEW model (squares, $k_1=1$), and the viscoelastic qEW model (circles, $k_r=0.1$, $k_c=0.9$). The bars indicate the value of the dispersion of $\sigma$, which is observed to tend to zero as $k_0\to 0$. Dotted lines are power law fittings of the form $\sigma(k_0)=\sigma(0)-\alpha k_0^{\gamma}$. The best fitted values of $\sigma(0)$ in the two cases are $\sigma_0=1.415$, and $\sigma_1=1.493$.
}
\label{sigma-de-k0}
\end{figure}

Here we concentrate in the one-dimensional case, that we have found behaves very differently to the higher dimensional cases discussed in \cite{jlr}. 
In Fig. \ref{sigma-de-k0} we show results for the stress across the surface in constant velocity driving. We indicate the average and the fluctuation of the stress across the interface, for a viscoelastic qEW model with spring constants $k_r=0.1$, $k_c=0.9$, and for a reference qEW model, with interface stiffness $k_1=k_r+k_c=1$. A fundamental fact is that for $k_0\to 0$, the limiting value for the viscoelastic qEW model is lower than for the standard qEW model. We note these two limiting values as $\sigma_0$ and $\sigma_1$, respectively. In addition, we see that the fluctuations of $\sigma$ across the whole interface (represented as bars in Fig. \ref{sigma-de-k0}) tend to zero as $k_0\to 0$ both for the standard, and the viscoelastic case, indicating a convergence to a constant force driving scenario in both cases (contrary to the higher dimensional viscoelastic qEW case).

On the basis of the results in Fig. \ref{sigma-de-k0} we will now describe the behavior of the viscoelastic model in a constant force driving situation. 
If $\sigma$ is sufficiently large, there is not any stationary solution to Eq. \ref{hdot}. 
This means that the system evolves by a single avalanche that lasts forever. The dynamics of this avalanche develops in a time scale of the order of $\eta$. In our time units in which $\eta \to 0$, we will consider this velocity as diverging, $v\to \infty$. As this dynamics is much more rapid than the viscoelastic one, the elements $\eta_u$ remain blocked during the time evolution. This means that in this regime the model behaves as a standard qEW model with an elastic constant $k_r+k_c$. The value $\sigma_1$ in Fig. \ref{sigma-de-k0} is precisely the depinning stress of this elastic model, so this regime occurs for all $\sigma>\sigma_1$.
In the opposite limit of small $\sigma$, namely $\sigma<\sigma_0$, the interface reaches a stationary configuration in which it remains pinned, and its velocity is zero.

\begin{figure}[h]
\includegraphics[width=8cm,clip=true]{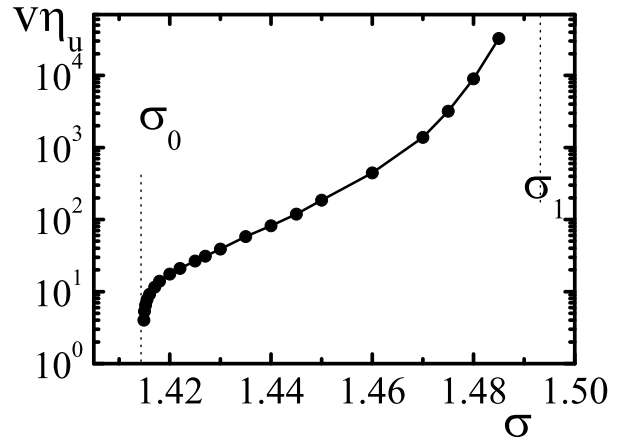}
\caption{Velocity as a function of stress in the creep regime. The vertical dotted lines indicate the values of $\sigma_0$ and $\sigma_1$
from the fits in the previous figure.}
\label{f2}
\end{figure}

The intermediate regime $\sigma_0<\sigma<\sigma_1$ is the viscoelastic creep regime in which we are mostly interested here. 
We present in Fig. \ref{f2} the results of straightforward numerical simulations in which a constant $\sigma$ is fixed, and the average velocity of the interface is measured. The vanishing of the velocity as $\sigma\to \sigma_0$ is clearly observed in this plot. 
In the range $\sigma_0<\sigma<\sigma_1$ the velocity of the interface remains finite, indicating that the interface does not reach any globally stable configuration. We stressed already that this is a property of a one dimensional interface. In fact, the results in \cite{jlr} on two-dimensional, and large dimensional systems indicate that such an intermediate regime does not exist in those cases.
As $\sigma\to\sigma_1$ we observe a divergence in the velocity. This divergence actually signals the transition from a velocity that is order $\eta_u^{-1}$ for $\sigma<\sigma_1$, to one of the order of $\eta^{-1}$ for $\sigma>\sigma_1$.
We are interested in characterizing in more detail this intermediate creep regime of the dynamics.

\begin{figure}[h]
\includegraphics[width=8cm,clip=true]{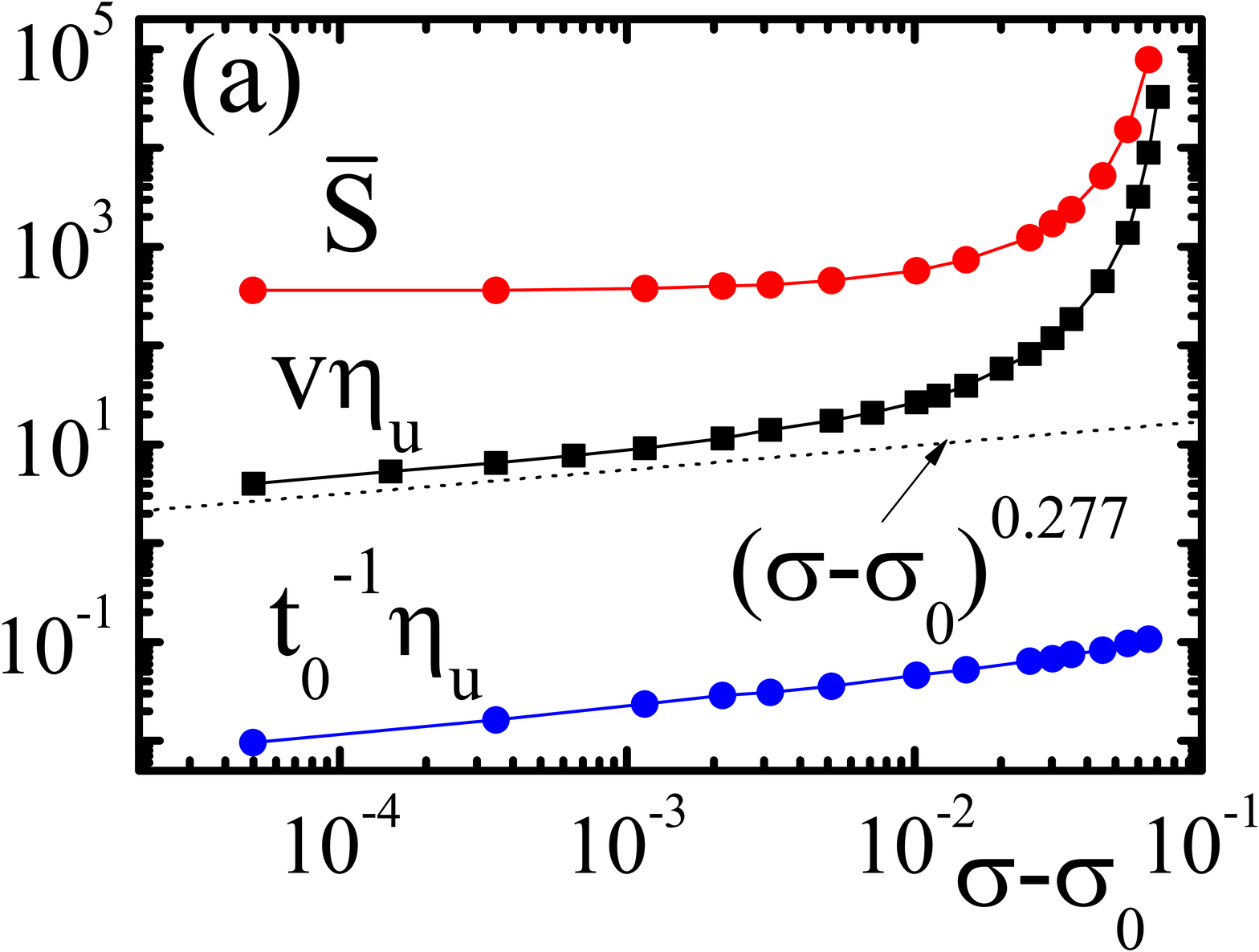}
\includegraphics[width=8cm,clip=true]{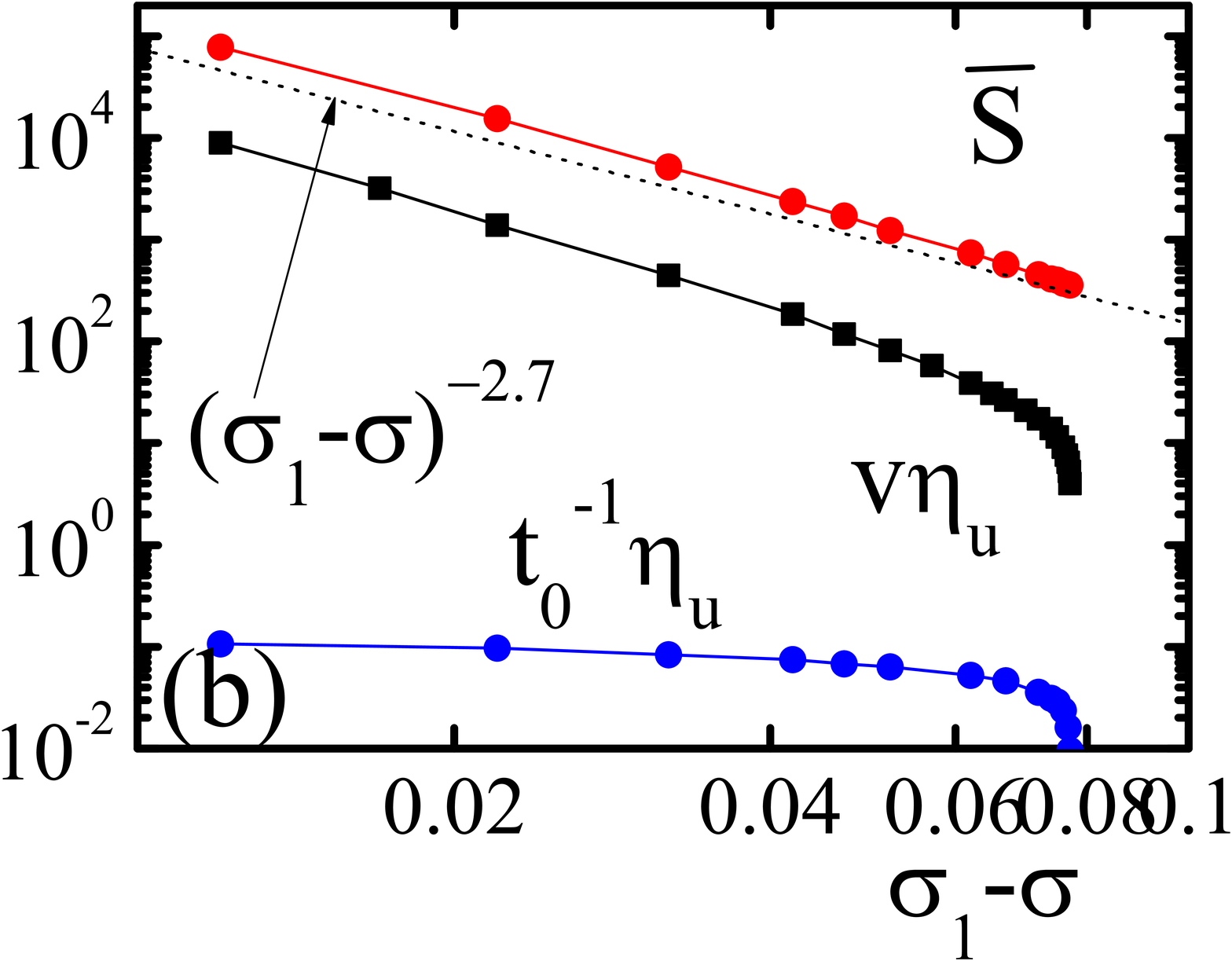}
\caption{Velocity as a function of stress and its decomposition as the ratio of an average avalanche size $\overline S$ and an inter-avalanche time $t_0$. The two plots highlight the power law dependencies for $\sigma\to\sigma_0$(a) and $\sigma\to\sigma_1$(b). In (a) the expected result of the velocity as $\sigma\to\sigma_0$ for a directed percolation process is indicated by the dashed line. In (b), the dashed line shows the expected behavior of $\overline S$ as $\sigma\to\sigma_1$ from the results of a qEW model with $k_1=k_r+k_c=1$. The values of $\sigma_0$ and $\sigma_1$ here are those obtained in Fig. \ref{sigma-de-k0}
}
\label{f3}
\end{figure}

The advance of the interface for $\sigma_0<\sigma<\sigma_1$ occurs through abrupt avalanches, that are instantaneous for our choice $\eta\to 0$. 
During avalanches the stretching of the dashpots remain fixed. Immediately after an avalanche, the dashpots are unrelaxed, an tend to equilibrium in a time scale of the order of $\eta_u^{-1}$. This relaxation triggers eventually new avalanches that maintain the interface in motion forever. This is the mechanism that generates a velocity of the order of $\eta_u$, in the creep regime $\sigma_0<\sigma<\sigma_1$.

Each avalanche can be characterized by the spatial coordinate $i$ at which it starts, its time of occurrence $t$ and its size $S$, that is defined as the sum of the  displacements of all sites that participate in it. An important quantity to consider is the size distribution of avalanches $N(S)$, such that $N(S)dS$
is the number of avalanches in the interval $S$, $S+dS$ per unit of time and unit of length in the system. The velocity of the interface can be written in terms of $N(S)$ as

\begin{equation}
v=\int  S N(S) dS.
\label{potencial}
\end{equation}
It is convenient to introduce an average time $t_0$ between avalanches (per unit of system length), such that $t_0=(\int  N(S)dS)^{-1}$. We can then write $v$ as 

\begin{equation}
v=\overline S/t_0
\end{equation}
where $\overline S \equiv \int  S N(S)dS/\int  N(S)dS$ is the average size of avalanches. In Fig. \ref{f3} we plot the results for $t_0$ and $\overline S$ from the numerical simulations. As we see, a divergence of $\overline S$ controls the divergence of velocity for $\sigma\to\sigma_1$, whereas a divergence of $t_0$ controls the vanishing of $v$ for $\sigma\to\sigma_0$.
\begin{figure}[h]
\includegraphics[width=8cm,clip=true]{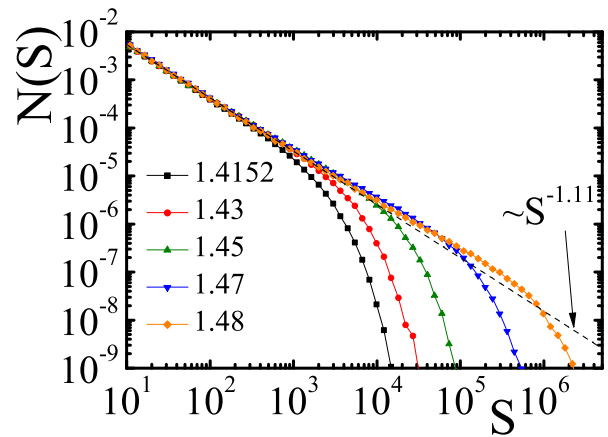}
\caption{The size distribution of avalanches in the creep regime, for different values of $\sigma$, as indicated.
}
\label{f4}
\end{figure}

\begin{figure}[h]
\includegraphics[width=8cm,clip=true]{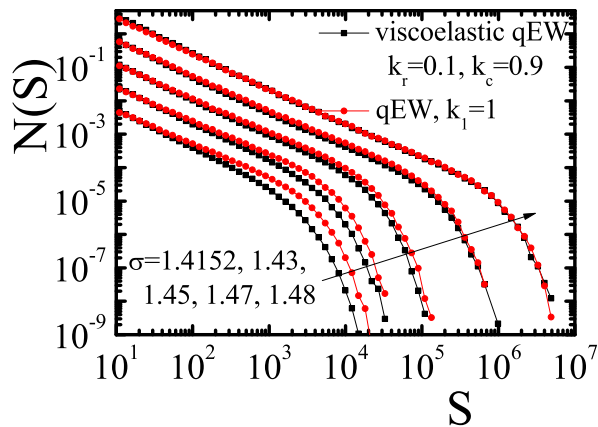}
\caption{Size distribution of avalanches in the viscoelastic qEW model in the creep regime, compared with the distribution of avalanches in a standard qEW model driven at constant force, in which the avalanches are randomly triggered (pairs of curves at different $\sigma$ have been vertically displaced for clarity). The distributions in the two cases tend to coincide as $\sigma\to\sigma_1\simeq 1.493$.
}
\label{compara}
\end{figure}

The size distribution of avalanches for different values of $\sigma$ is presented in Fig. \ref{f4}. We observe the development of a critical distribution as $\sigma\to\sigma_1$, with a decaying exponent $\tau\simeq 1.11$, which is similar to that corresponding to the qEW model. We have here a clear indication that the accelerated creep as $\sigma\to \sigma_1$ is caused by the existence of large avalanches in this regime.
An examination of the epicenters of the avalanches in the present case, reveals that they are not temporally nor spatially correlated.
This seems reasonable in this regime: if one large avalanche is triggered, relaxation may produce a subsequent avalanche essentially in any point affected by the first one, but not necessarily close to the epicenter of the first one.
In addition, as large avalanches develop as in a standard qEW model with stiffness $k_r+k_c$ this allows to understand the size distribution obtained in the following way: the avalanches in the limit $\sigma\to\sigma_1$ are equivalent to those of a normal qEW model with constant force driving, in the case in which avalanches are triggered in random positions of the system. In fact, a direct comparison of the two cases (Fig. \ref{compara}) confirms this equivalence. 

For the qEW model, the divergence of $\overline S$ as 
$\sigma\to\sigma_1$ is given by $\overline S \sim S_{max}^{2-\tau}\sim(\sigma_1-\sigma)^{(2-\tau)(1+\zeta)}$, with $\zeta$ being the roughness exponent. We obtain $\overline S \sim (\sigma_1-\sigma)^{2.7}$. This dependence of $\overline S$ on $\sigma\to\sigma_1$ controls the divergence of $v$, as the time between avalanches $t_0$ becomes constant in this limit.  In Fig. \ref{f3}(b) we can see in fact that the behavior of the creep velocity is compatible with this analysis.

\begin{figure}[h]
\includegraphics[width=8cm,clip=true]{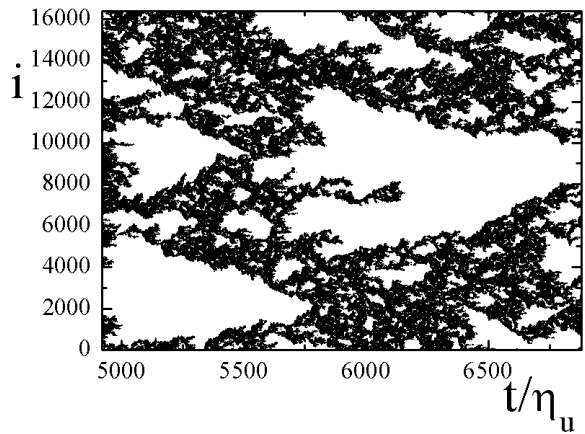}
\caption{Epicenters of the events across the system as a function of time, for $\sigma=1.4152$. The characteristics of a contact process, in which new events are activated by previous nearby ones is apparent. 
}
\label{f5}
\end{figure}

When $\sigma$ is reduced towards $\sigma_0$, the mean size of avalanches remains finite, as Fig. \ref{f3}(a) shows.
This implies that spatial correlations between consecutive avalanches must be observed, as the relaxation of the dashpots can only trigger new avalanches in the close vicinity of regions affected by a previous one.
In addition, the vanishing of the creep velocity
as $\sigma\to\sigma_0$ implies a divergence of the average inter-avalanche time $t_0$. It is interesting to look at the spatial distribution of avalanches to see how this happens. 
We thus run a simulation for $\sigma$ close to $\sigma_0$, and once a stationary creep situation is achieved, plot the location of the epicenters of every event in the system,  as a function of time. The result, presented in Fig. \ref{f5}, reveal a striking spatial structure. 
Those parts of the system that are active at some particular time, continue to trigger avalanches in neighboring places, at a non-singular rate.
There are also large spatio-temporal regions that are free of avalanches. 
As $\sigma$ is reduced, it is observed that the parts of the system that remain active are more scarce
making the average inter-avalanche time increase.  

These findings clarify the way in which creep velocity vanishes as $\sigma_0$ is approached. 
The structure in Fig. \ref{f5} reveals the existence of a {\em contact process}, in which a given avalanche can activate posterior ones within a time interval of the order of $\eta_u$, and no further away than the maximum extension of avalanches (which is limited to about 200 lattice sites for the parameters used here). A well known conjecture by Janssen and Grassberger
\cite{j-g} suggests that this kind of contact process must in general belong to the universality class of directed percolation.
If this is the case, the velocity of the interface in the limit $\sigma\to\sigma_0$ (which is proportional to the density of events in Fig. \ref{f5}) must follow a law $v\sim (\sigma-\sigma_0)^\beta$, with $\beta\simeq 0.277$. Our results are consistent with this dependence (see Fig. \ref{f3}(a)). 

\section{Discussion and conclusions}

In this paper we have analyzed the possibility to describe a fatigue crack growth situation at constant applied load  by modeling the advance of the crack edge through a viscoelastic qEW model. In the absence of any thermally activated effects, we have seen that three clearly separated regimes appear as a function of the applied load: no crack advance if $\sigma<\sigma_0$, unstable crack advance (that can be described as immediate breaking of the system)  if $\sigma>\sigma_1$, and a creep regime if $\sigma_0<\sigma<\sigma_1$, where the crack velocity is controlled by a viscoelastic coefficient $\eta_u$. 

We have analyzed in detail the dynamics of the model in the creep regime. 
Our model displays naturally a fatigue limit $\sigma_0$ below which the advance of the crack is completely halted.
As the stress is diminished towards the fatigue limit $\sigma_0$, the velocity vanishes as a power law. 
We have interpreted this behavior in terms of a contact process in which one avalanche can give rise to successive ones within a limited spatial range, and in a typical time controlled by $\eta_u$.
As $\sigma\to\sigma_1$ the velocity diverges (in the scale of $\eta_u$) as a power law, until the unstable crack growth regime sets in for $\sigma>\sigma_1$

Compared with the thermal creep regime studied in related models \cite{busting}, the most striking difference of the model studied here is the existence of a fatigue limit, corresponding to a stress below which the time to failure of the system is truly infinite. This behavior, which has been observed experimentally in different materials\cite{fatiga1} has been explained before relying in {\em ad hoc} healing mechanisms \cite{kun1,kun2,alava}. Our model provides an alternative possible explanation for this phenomenon that appears exclusively because of the microscopic dynamics of the model. The presented mechanism of visco-elastic creep may thus serve to improve our understanding of failure mechanisms of solids under constant stress.

\section{Acknowledgments}

The author is financially supported by CONICET (Argentina). Partial support from grant PIP 112-2009-0100051 (CONICET, Argentina) is also acknowledged.


\begin{thebibliography}{4}

\bibitem{ferro1} F. Colaiori, Adv. Phys. {\bf 57}, 287 (2008).
\bibitem{ferro2} S. Zapperi, P. Cizeau, G. Durin, and H. E. Stanley, Phys. Rev.  B {\bf 58}, 6353 (1998).
\bibitem{ferro3} S. Lemerle, J. Ferr\'e, C. Chappert, V. Mathet, T. Giamarchi and
P. Le Doussal, Phys. Rev. Lett. {\bf 80}, 849 (1998).
\bibitem{wet1}E. Rolley, C. Guthmann, R. Gombrowicz, and V. Repain, Phys.
Rev. Lett. {\bf 80}, 2865 (1998).
\bibitem{wet2}P. Le Doussal, K. J. Wiese, S. Moulinet, and E. Rolley,
Europhys. Lett. {\bf 87}, 56001 (2009).
\bibitem{eq1} Y. Ben-Zion and J. R. Rice, J. Geophys. Res. {\bf 98}, 14109 (1993).
\bibitem{eq2} D. S. Fisher, K. Dahmen, S. Ramanathan, and Y. Ben-Zion,
Phys. Rev. Lett. {\bf 78}, 4885 (1997).
\bibitem{eq3} D. S. Fisher, Phys. Rep. {\bf 301}, 113 (1998).
\bibitem{depinning} D. S. Fisher, Phys. Rev. B {\bf 31}, 1396 (1985).
\bibitem{creep1} T. Nattermann and S. Scheidl, Adv. Phys. {\bf 49}, 607 (2000).
\bibitem{creep2} P. Chauve, T. Giamarchi, and P. Le Doussal, Phys. Rev. B {\bf 62}, 6241 (2000).
\bibitem{fatiga1} S. Suresh, {\it Fatigue of Materials} (Cambridge University
Press, Cambridge, U.K., 1998), 2nd ed.
\bibitem{fracture1}M. Alava, P. K. Nukala, and S. Zapperi, Adv. Phys. {\bf 55}, 349 (2006).
\bibitem{fatiga2} We refer exclusively here to {\em static} fatigue failure, that occurs due to a long time application of a {\em constant} load. It must not be confused with {\em cyclic} fatigue processes, in which the failure occurs due to the interplay of oscillating loads and plastic effects in the material.
\bibitem{basquin}O. H. Basquin, {\em Proceedings of American Society of Testing
Materials ASTEA} (10), 625 (1910).
\bibitem{kun1}F. Kun, H. A. Carmona, J. S. Andrade, Jr., and H. J. Herrmann, Phys. Rev. Lett. {\bf 100}, 094301 (2008).
\bibitem{kun2}F. Kun, M. H. Costa, R. N. Costa Filho, J. S. Andrade Jr, J. B. Soares, S. Zapperi and H. J. Herrmann, 
J. Stat. Mech. (2007) P02003.
\bibitem{alava}M. J. Alava, J. Stat. Mech. (2007) N04001.
\bibitem{hinri}H. Hinrichsen Adv. Phys. {\bf 49}, 815 (2000).
\bibitem{stanley}A. Barabasi and H. Stanley, {\em Fractal Concepts in Surface Growth} (Cambridge University Press, Cambridge, 1995). 
\bibitem{jlr}E. A. Jagla, F. P Landes, and A. Rosso, arXiv:1310.5051.
\bibitem{j-g}H. K. Janssen, Z. Phys. B {\bf 42}, 151 (1981);
P. Grassberger, Z. Phys. B {\bf 47}, 365 (1982).
\bibitem{busting}S. Bustingorry, A. B. Kolton, and T. Giamarchi, Phys. Rev. E {\bf 85}, 021144 (2012).



\end{thebibliography}
\end{document}